\documentclass[aps,reprint,pre,linenumbers]{revtex4-1}
\usepackage[utf8]{inputenc}
\usepackage{mathtools}

\usepackage{graphicx}
\usepackage{dcolumn}
\usepackage{bm}
\usepackage{xcolor}
\usepackage{amsmath}
\usepackage{amssymb}
\usepackage{hyperref}

\newcommand{\ER}{Erd\H{o}s--R\' enyi{ }}

\begin{document}
\nolinenumbers
\title{How a minority can win: \\ Undemocratic outcomes in a simple model of voter turnout}

\date{\today}
\author{Ekaterina Landgren}
\email{ek672@cornell.edu}
\affiliation{Center for Applied Mathematics, Cornell University, Ithaca, New York 14853, USA}
\author{Jonas L. Juul}
\affiliation{Center for Applied Mathematics, Cornell University, Ithaca, New York 14853, USA}
\author{Steven H. Strogatz}
\affiliation{Center for Applied Mathematics, Cornell University, Ithaca, New York 14853, USA}

\begin{abstract}
The outcome of an election depends not only on which candidate is more popular, but also on how many of their voters actually turn out to vote. Here we consider a simple model in which voters abstain from voting if they think their vote would not matter. Specifically, they do not vote if they feel sure their preferred candidate will win anyway (a condition we call complacency), or if they feel sure their candidate will lose anyway (a condition we call dejectedness). The voters reach these decisions based on a myopic assessment of their local network, which they take as a proxy for the entire electorate: voters know which candidate their neighbors prefer and they assume---perhaps incorrectly---that those neighbors will turn out to vote, so they themselves cast a vote if and only if it would produce a tie or a win for their preferred candidate in their local neighborhood.
We explore various network structures and distributions of voter preferences and find that certain structures and parameter regimes favor undemocratic outcomes where a minority faction wins, especially when the locally preferred candidate is not representative of the electorate as a whole.
\end{abstract}

\maketitle

\section{Introduction} \label{sec:intro}

Election forecasting is a difficult problem with real-world consequences~\cite{wang2015forecasting,rothschild2009forecasting,volkening2020forecasting}. Part of the difficulty is that human psychology is murky. How do voters decide which candidate they prefer? What makes them change their minds? And how do they decide whether to tell pollsters what they really think? More broadly,  modeling elections and voter behavior can shed light on a wide range of puzzling issues about human decision-making and hot-topic phenomena such as polarization and the formation of political echo chambers~\cite{baumann2020modeling,madsen2018large,hayat2017you,heider1946attitudes,kulakowski2005heider,marvel2011continuous,de2020emergence,del2017modeling}.

There is a rich literature on agent-based opinion dynamics.  This literature includes the ``voter model'' of probability theory~\cite{holley1975ergodic} and its many extensions (see \cite{redner2019reality} for a review), as well as bounded confidence models~\cite{lorenz2007continuous,hegselmann2002opinion, hickok2021bounded}. In such models, agents interact on a network and change their opinions according to certain rules.
For example, the agents can adopt the opinion of one of their nearest neighbors chosen at random~\cite{holley1975ergodic}, or they can adopt the opinion held by the majority of their  neighbors~\cite{krapivsky2003dynamics,galam2008sociophysics}, or they can update their opinion at a nonlinear rate depending on the opinion distributions of their neighbors~\cite{lambiotte2007dynamics,lambiotte2008dynamics,slanina2008some}. The update rules can also depend on the state of agents' opinions (e.g., introducing stubborn \cite{masuda2010} or confident \cite{volovik2012dynamics} voters who do not change their opinions easily). A key question is when consensus forms among the nodes and what conditions promote it.

However, opinion dynamics is just one facet of voter behavior. In the real world, another important factor is voter turnout, defined as the percentage of eligible voters who cast a ballot in an election.
The turnout rate depends on many socioeconomic, political, and institutional factors, from population size to campaign expenditures to registration requirements~\cite{geys2006explaining, cancela2016explaining}. The abundance of relevant factors makes predicting voter turnout difficult.

One factor influencing voter turnout is the closeness of the election~\cite{endersby2002closeness, bursztyn2017polls}. Intuitively, one might expect that close elections should produce higher turnout, but some scholars dispute that this is the case~\cite{matsusaka1993election, gerber2020one}. Here we explore the effect of network structure on individual agents' perceptions of election closeness and the consequent impact on turnout and on the election itself. 

Certain network structures and opinion distributions can lead to minority nodes mistakenly believing that they belong to a majority. The phenomenon whereby local knowledge of the network is not representative of the electorate as a whole is known as the ``majority illusion''; a ``minority illusion'' is also possible~\cite{lerman2016majority}. We are interested in conditions that allow a minority to win elections by generating a higher turnout than the majority. 

The undemocratic phenomenon of the minority defeating the majority has been studied previously in many ways. For example, Iacopini et al. \cite{iacopini2021vanishing} examine when a minority can build a critical mass to cause a cascade on hypergraphs and become the dominant opinion. In a similar spirit, Touboul~\cite{touboul2019hipster} and Juul and Porter~\cite{juul2019hipsters} examine how antiestablishment nodes (nodes that prefer to belong to a minority) can spread their influence and create an antiestablishment majority.


 In this paper we consider a model of voter turnout that allows for majority and minority illusions. We ask: What network structures enable minority factions to win? While we do not consider opinion dynamics (our model voters never change their minds), the mechanisms of voter turnout alone can generate situations where a small minority can win in a landslide. This counterintuitive result is one of our main findings. Whether it holds in more realistic models remains to be seen.

The paper is laid out as follows. Section \ref{sec:model} introduces the model. In Section \ref{sec:results}, we apply the model on a variety of network structures: Erd\H{o}s-R\'enyi networks (\ref{sec:ER}), stochastic block networks (\ref{sec:SB}), scale-free networks (\ref{sec:scale-free}), and random geometric networks (\ref{sec:geo-rand}). Section~\ref{sec:discussion} summarizes and discusses the results.

\begin{figure*}[!tb]
    \centering
    \includegraphics[width=.9\linewidth]{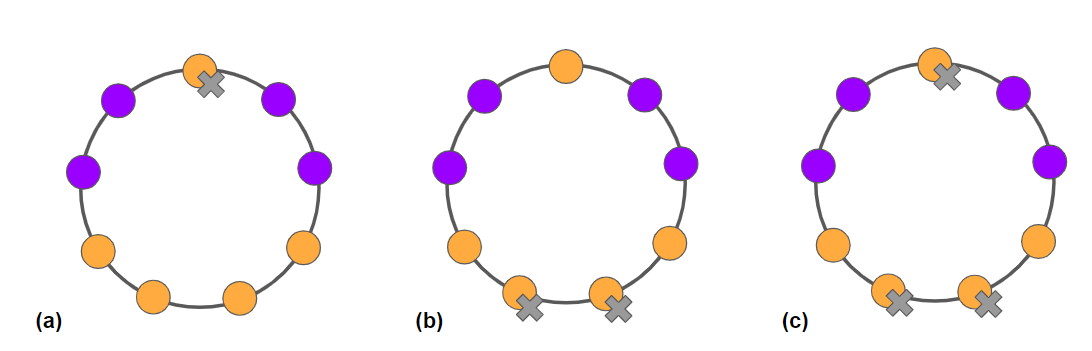}
    \caption{ The behavioral assumptions of  (a) dejectedness, (b) complacency, and (c) their combination applied to the same ring network with a $5-4$ split between orange and purple nodes. (a) The top orange node is  surrounded by a purple node on either side. Thus, in its local (one-hop) neighborhood it is outnumbered $2-1$, so its vote cannot tie or win the upcoming election \emph{in that local neighborhood}.  Making a myopic (and wrong) estimate of the orange opinion's chances globally, based solely on its local neighborhood, the orange node believes its vote cannot affect the upcoming election, so it gets dejected and does not cast a vote, as indicated by the gray cross. (b) The two bottom nodes are completely surrounded by other orange nodes. Based on this local information, they erroneously conclude that the upcoming election is a safe win, become complacent, and do not vote. (c) Because three orange nodes do not vote, purple wins the overall election by $4-2.$}
  \label{fig:ring_graphs}
\end{figure*}

\section{The model} \label{sec:model}
Our simplified model of voter behavior is intended to spotlight the role of two social effects: \textit{complacency} and \textit{dejectedness}. In the model, voters have fixed opinions and only need to decide whether to participate in an upcoming election. Whether a node chooses to vote or abstain depends on whether its local neighborhood causes the node to experience complacency, dejectedness, or neither of these effects.  Complacency is the effect where nodes that are surrounded predominantly by nodes with matching opinions do not bother to vote, because they are convinced that their preferred candidate is going to win in any case. Dejectedness is the effect where nodes that are surrounded predominantly by nodes with opposite opinions tend not to vote, because they are convinced that the situation is hopeless and their preferred candidate is going to lose. 

Our model of voter behavior under dejectedness and complacency can be introduced formally as follows. We assume that $N$ voters live on a network, and each node has some opinion $\theta$, drawn from a probability distribution $f(\theta)$. We shall assume that only two opinions exist, although studying the more general case of multiple opinions is a natural direction for future work. In the context of the model, these opinions can be thought of as preferences for one of two candidates in an election, but they could also represent binary referendum options, or any other binary choice. 

Continuing in the spirit of simplicity, we further assume that each node knows the opinion of all its neighbors. The only question is who will vote. Whether a node decides to vote or not depends on whether it thinks its vote will make a difference, which in turn depends on the prevalence of the two opinions among its neighbors in the network. We assume the following simple-minded \emph{decision rule}: A node chooses to cast its ballot if and only if its vote would cause a tie or a one-vote win in its one-hop network neighborhood (assuming that all its neighbors choose to vote). More precisely, if a focal node with opinion $\theta$ has $k_{\theta}$ neighbors with opinion $\theta$ and $k_{\phi}$ neighbors with the opposite opinion $\phi$, it will vote if and only if 
$$0\leq k_{\phi}-k_{\theta} \leq 1.$$


Figure \ref{fig:ring_graphs} illustrates the model. In the example shown, nine nodes live on a ring graph. Five nodes hold a majority opinion (orange) and four nodes hold a minority opinion (purple). Figure~\ref{fig:ring_graphs}(a) illustrates the effect of dejectedness. When the top orange node decides whether to cast its vote, it sees that both of its neighbors hold the opposite opinion. Since purple outnumbers orange in the top node's neighborhood, even if the orange node decides to vote it cannot tie or win the election locally, so it gets dejected and abstains from voting (as indicated by the gray cross). Figure~\ref{fig:ring_graphs}(b) illustrates the effect of complacency. The two orange nodes at the bottom are completely surrounded by nodes with the same orange opinion. These two nodes conclude that orange is a local majority, even without their votes, and thus abstain from voting. Figure~\ref{fig:ring_graphs}(c) shows the result of the election: $3$ orange nodes abstain from voting, leading to a $4-2$ win by the purple minority. 

As this example shows, the election outcome depends on a surprisingly subtle interplay among three factors: the network structure, the proportion of nodes that hold each of the two opinions, and how the opinions are arranged among the network nodes. Thus, this result raises several questions: Are some network structures more likely to result in minority wins than others, at least under our model? Does homophily (the tendency for neighboring nodes to hold identical opinions) increase or decrease the likelihood of minority wins? And how does the minority size affect the likelihood of a minority win? In the remainder of this paper, we pursue these questions by simulating our model on various network topologies, and with different choices for the arrangement of opinions among the network nodes.

\section{Model Networks} \label{sec:results}

Our model networks (``electorates'') consist of  $N$ nodes (``voters''), of which $N_{+}$  hold the majority opinion and $N_{-}$ hold the minority opinion. We typically work with networks of size $N=100$, in which  case $N_{-}$ can also be interpreted as the \emph{minority fraction}, defined as the percentage of the electorate that holds the minority opinion. For each class of networks, we treat $N_{-}$ as a control parameter and explore how the probability of a minority victory depends on $N_{-}$. In our analytical work on stochastic block networks (Section \ref{sec:SB}), we also find it convenient to express the results as a function of the ratio
$$\alpha =\frac{N_{-}}{N_{+}} \le 1,$$ 
a parameter that quantifies how closely divided the electorate is.  

\subsection{Erd\H{o}s-R\'enyi networks\label{sec:ER}}
We begin by applying our model to Erd\H{o}s-R\'enyi random graphs~\cite{newman_networks_2018}.
In these networks, any given pair of nodes is connected by an undirected edge with  probability $p$. 
Since the number of nodes, $N$, and the edge probability, $p$, define this family of random graphs, the family is often denoted $G(N,p)$.  

\begin{figure}[tb]
    \centering
    \includegraphics{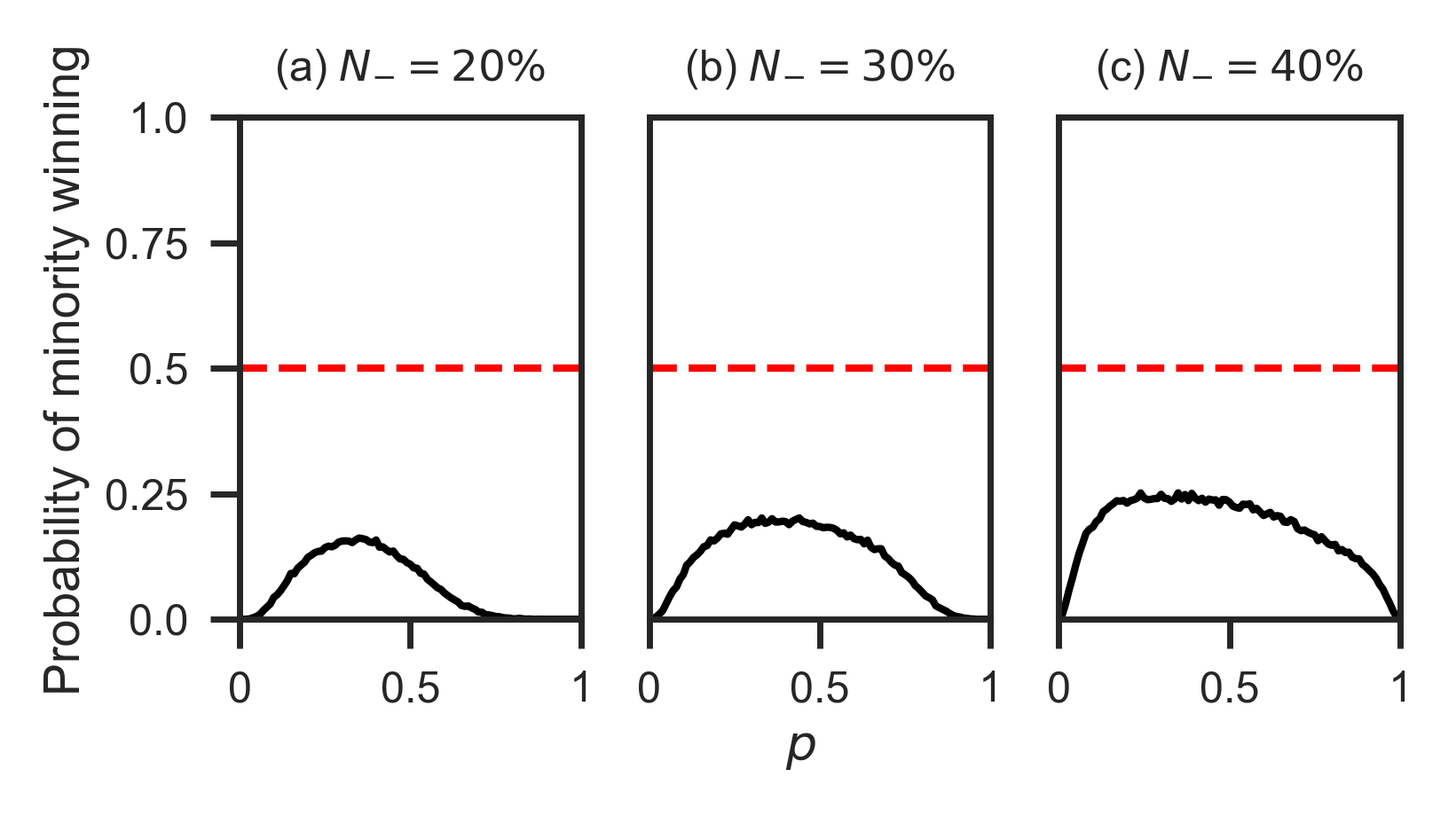}
    \caption{
  Undemocratic outcomes are rare on Erd\H{o}s-R\'enyi random graphs. The plots show the proportion of minority victories on random graphs drawn from the family $G(100,p)$ as a function of the edge probability $p$ ($x$-axis) for three different  scenarios: (a) a minority that is 20\% of the entire electorate, (b) 30\% minority, and (c) 40\% minority. Each data point in a plot is based on $10^6$ numerical experiments. For all three scenarios, the peak probability of a minority victory occurs for an intermediate $p$. But note that the minority never wins more than half of the time; the curves lie below the dashed red line at all values of the edge probability $p$.} 
  \label{fig:ER}
\end{figure}

Figure~\ref{fig:ER} shows how the average proportion of undemocratic outcomes changes as we vary the edge probability $p$, for fixed network size $N=100$ and three different choices for the minority fraction $N_{-}$. 
In Fig.~\ref{fig:ER}(a), the majority nodes outnumber the minority nodes by $80$ to $20$, a considerable margin. Under these circumstances it is not easy for the minority to pull off an upset win, but it is possible, thanks to the complacency of the majority. The probability that minority wins peaks at around $p=0.25$, with a corresponding win probability of less than $0.2$. Figure~\ref{fig:ER}(b) shows the corresponding plot when we increase the fraction of minority nodes to $30$ out of $100$, and Fig.~\ref{fig:ER}(c) does the same for $40$ minority nodes. The effects of these changes are mild. The main things to notice are that as the electorate becomes more nearly evenly split, the peak probability that the minority wins becomes slightly higher
and there is a widening of the range of $p$-values where minority wins occasionally take place. Still, the main message of Fig.~\ref{fig:ER} is  that undemocratic outcomes are fairly rare on this class of random graphs. Indeed, in our simulations of the model on Erd\H{o}s-R\'enyi networks, there is no parameter regime where a minority wins most of the time.  

From Figure~\ref{fig:ER}, we can make two  observations about when a minority can win: minority victories become more likely for larger minorities and for intermediate values of $p$. The first observation makes intuitive sense: A minority victory is less likely when the margin between the number of majority nodes and the number of minority nodes is wider, because fewer minority nodes means that more majority nodes must abstain from voting in order to ensure a minority win.
Second, to understand why minority wins are most likely for intermediate values of $p$, it is helpful to consider the extreme network structures that can arise in Erd\H{o}s-R\'enyi networks. There are two such extremes. When $p=0$, the network has $N$ components, each consisting of a single node, and no node has neighbors. In the absence of local information, every node votes, making undemocratic outcomes impossible. At the other extreme, when $p=1$ the Erd\H{o}s-R\'enyi network becomes a complete graph. On a complete graph, every node has perfect information about the global state of the network, which leads to dejectedness for the minority nodes and complacency for the majority nodes (if the margin is greater than 1). As long as this condition holds true, nobody votes, and therefore undemocratic outcomes do not occur in this case either. 

\subsection{Stochastic block networks} \label{sec:SB}
In section \ref{sec:ER}, we assumed that opinions were distributed uniformly at random among the nodes in \ER networks. 
Distributing the opinions in this way meant that there was no homophily in the networks. 
Looking back at Fig.~\ref{fig:ring_graphs}, we see that the nodes that are resistant to complacency and dejectedness have the minority and minority opinions nearly equally represented in their local neighborhoods. As such, it is the \emph{other} nodes, the ones in homophilous neighborhoods, that tend not to vote and thereby open the door to undemocratic outcomes. 
In other words, we expect homophily to play an important role in enabling the minority to win.

One way to introduce such homophily into randomly generated networks is to create random networks with community structure and assume that nodes in the same community have the same opinion. 
We now do exactly this by simulating our model on ``stochastic block networks''~\cite{newman_networks_2018}.

It is helpful to think of stochastic block networks as a generalization of Erd\H{o}s-R\'enyi networks. Whereas in Erd\H{o}s-R\'enyi networks, the probability of forming an edge is the same for any two pairs of nodes, in stochastic block networks the node set is partitioned into disjoint subsets. 
The probability of forming an edge between nodes then depends on the nodes' respective subsets.
If the nodes are in the same subset, they are part of the same community, and the probability of them being joined by an edge is high. 
On the other hand, nodes in different subsets are assumed to not be part of the same community, and the probability of an edge between them is low. 

Since we are interested in the interactions between majority and minority nodes, we will use a stochastic block network with two blocks. The probability of forming an edge between two nodes can be represented as a matrix:
\begin{equation}
    P= 
    \begin{pmatrix}
    p_{11} &  p_{12} \\
    p_{21} & p_{22}
    \end{pmatrix},
\end{equation} 
where $p_{ij}$ is the probability of forming an edge for any pair of nodes from block $i$ and block $j$. For the sake of simplicity, we pick one in-block probability ($p_{11}=p_{22}=p_{\text{in}}$) and one inter-block probability ($p_{12}=p_{21}=p_{\text{out}}$) to reflect the in-group/out-group differences. These relative probabilities serve as a homophily parameter. When $p_{\text{in}}/p_{\text{out}}$ is high, the network exhibits high homophily, since nodes are more likely to form edges within their block. When $p_{\text{in}}/p_{\text{out}}$ is low, the network exhibits low, or even anti-homophily, since the nodes are more likely to form edges across blocks. In the special case $p_{\text{in}}=p_{\text{out}}$, we obtain Erd\H{o}s-R\'enyi networks.

\begin{figure}[!ht]
    \centering
    \includegraphics{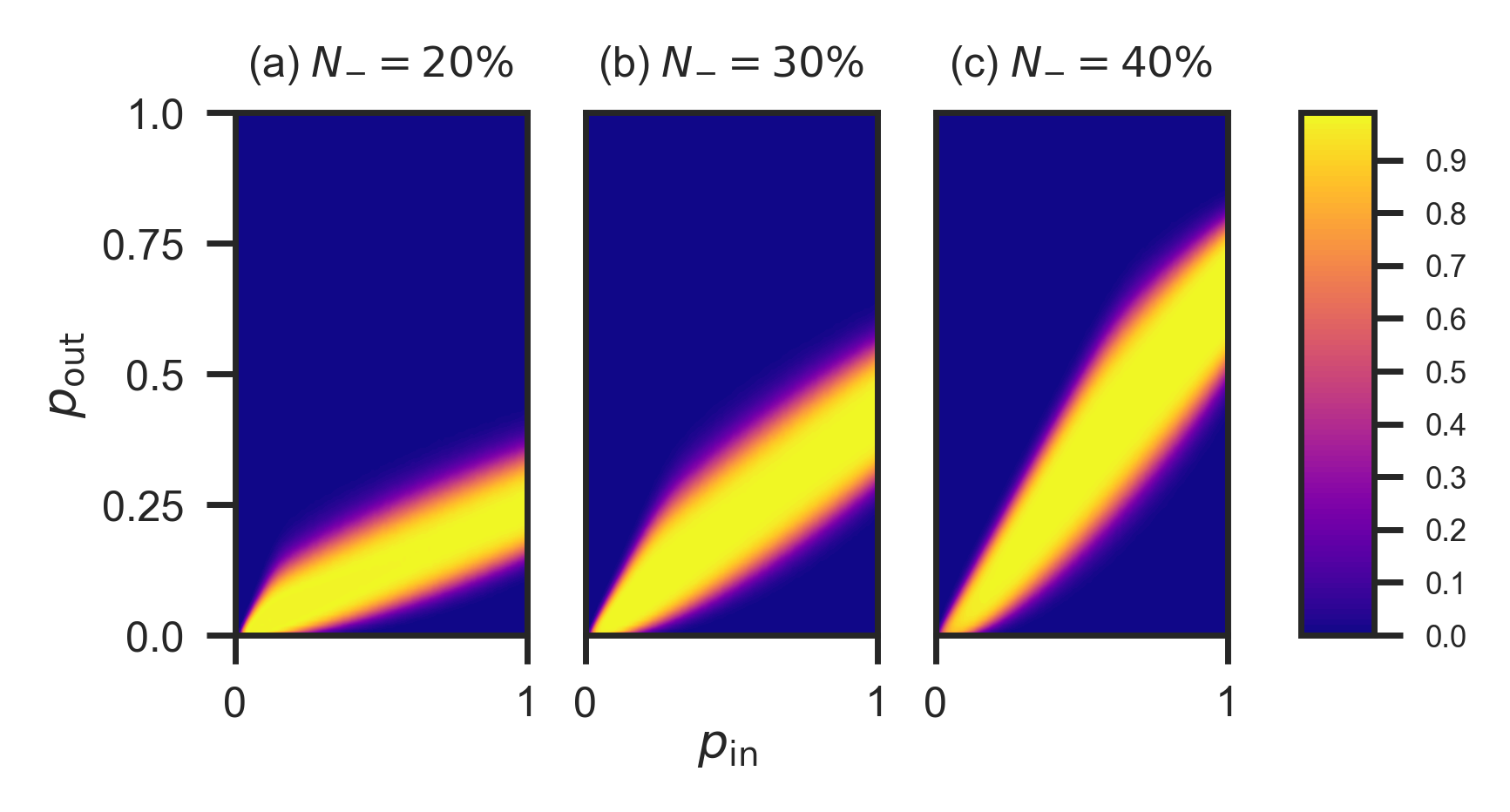}
    
    \caption{ Introducing community structure to random graphs allows for a prevalence of undemocratic outcomes within some parameter regions (the diagonal yellow regions). The proportion of undemocratic outcomes is shown in color as a function of $p_{\text{in}}$ and $p_{\text{out}}$ on stochastic block networks of $N=100$ nodes with minority blocks of sizes (a) $N_{-}=20$, (b) 30, and (c) 40 nodes, for $10^6$ simulations.} 
  \label{fig:SB}
\end{figure}


\subsubsection{Numerical experiments on stochastic block networks}

Figure \ref{fig:SB} shows the proportion of undemocratic outcomes as a function of $p_{\text{in}}$ and $p_{\text{out}}$ for networks where majority nodes outnumber minority nodes by varying amounts. The color represents the proportion of simulations in which the minority wins. Parameter values leading to undemocratic outcomes are conspicuous as the bright yellow regions. 


While there are quantitative differences among the three networks, there are important qualitative similarities. In each of the three panels, most of the parameter space is colored dark blue, corresponding to the democratic outcomes one would naturally expect. However, there are also yellow diagonal regions in which the minority wins more than half of the time. The highest probability of a minority victory occurs close to the midline of the yellow region, where $p_{\text{out}}/ p_{\text{in}}\approx \alpha$. While not visible in the figure, the global maximum occurs on the right edge of each panel, at the point where when $p_{\text{in}}=1$ and $p_{\text{out}} = \alpha.$
As we increase the size $N_{-}$ of the minority population, the location of the peak  moves up the $p_{\text{out}}$ axis, resulting in an increased slope of the yellow region, while $p_{\text{in}}$ stays pinned at its maximum value, $p_{\text{in}}=1$. 

These results confirm our intuition from the \ER networks: Undemocratic outcomes occur in the intermediate information regime. They do not thrive on complete networks, nor on fragmented ones with many components. Rather, they favor regimes where nodes have an intermediate level of knowledge about the state of the electorate as a whole. 
 
 An intuitive way of understanding Figure \ref{fig:SB} is to think about the effects of complacency and dejectedness. In order to avoid these effects, it is necessary to have both majority and minority opinion nearly equally represented in a node's neighborhood. Because there are more majority nodes in the network, at high $p_{\text{in}}$ and intermediate $p_{\text{out}}$ settings the minority nodes are most likely to know almost equal numbers of nodes who agree and disagree with them. However, in that same setting, the majority nodes are more likely to know more nodes who agree with them because of the high $p_{\text{in}}$ probability, and therefore are more likely to get complacent. This effect is what allows the minority to win.


 \subsubsection{Analytical results for stochastic block networks with $p_{\text{in}}=1$: Exact probability of a minority victory}
 
 For the convenient special case where $p_{\text{in}}=1$, we can find the probability of a minority victory exactly. To do so, observe that if the number of majority nodes exceeds the number of minority nodes by at least two ($N_{+}\geq N_{-}+2$), then none of the majority nodes will vote, due to the effects of complacency. Therefore, in this particular case, the minority will win as long as \emph{any}  minority node votes. We can compute the probability of that event in a few easy steps as follows.
 
The first step is to consider the probability that any \emph{given} minority node votes. Because $p_{\text{in}}=1$, the given minority node is certain to be linked to all the other minority nodes in the electorate and hence is sure to see exactly $N_{-}$ votes for the minority opinion in its local neighborhood (including its own vote). Now invoke the decision rule: the given minority node votes if and only if doing so would either cause a tie or a one-vote victory in its local neighborhood. For those events to happen, the minority node also needs to be connected to either the same number, $N_{-}$, of \emph{majority} nodes, or one less than that number. Those two events both happen according to binomial probability distributions, because they involve choosing either $N_{-}$ or $N_{-}-1$ majority nodes out of a total of $N_{+}$  available. Therefore, the probability that the given minority node votes is a sum of two binomial terms:
 \begin{equation}
 \begin{aligned}
     & P(\text{any given minority node votes}) \\ &={N_{+} \choose N_{-}} p_{\text{out}}^{N_{-}} \left(1-p_{\text{out}}\right)^{N_{+}-N_{-}} \\
     &+{N_{+} \choose N_{-}-1} p_{\text{out}}^{N_{-}-1} \left(1-p_{\text{out}}\right)^{N_{+}-(N_{-}-1)}.
      \label{eq:step1}
     \end{aligned}
 \end{equation}
 The first term expresses the probability that a minority node sees an equal number of majority and minority nodes (and will vote because it can cause a local tie). The second term represents the probability that the minority node sees $N_{-}-1$ majority nodes (and will vote because it can cause a local minority victory). All other possibilities are irrelevant: If the minority node sees more than $N_{-}$ majority nodes, it would become dejected, whereas if it sees fewer than $N_{-}-1$ majority nodes, it would become complacent.

\begin{figure}[h]
    \centering
    \includegraphics{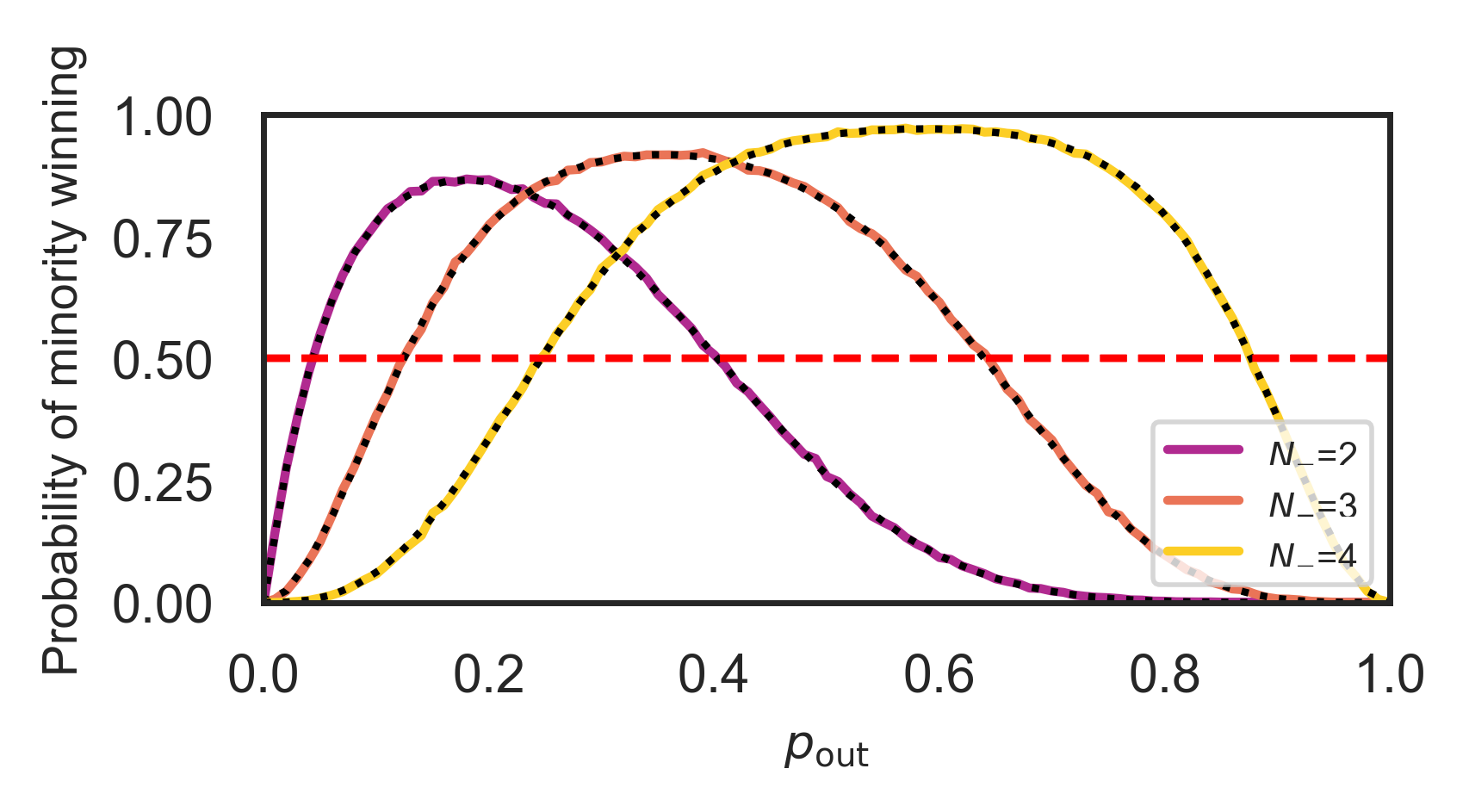}
    \caption{The proportion of undemocratic outcomes on stochastic block networks for the special case $p_{\text{in}}=1.$ The probability of a minority victory is plotted as a function of $p_{\text{out}}$, for networks of size $N=10$. Results for three values of  $N_{-}$ are shown, corresponding to minority fractions of $20\%$ (red), $30\%$ (orange), and $40\%$ (yellow). The dotted black lines show the analytical expression in Eq.~\eqref{eq:all-alpha-maintext}, which agrees with numerical results from $10^6$ simulations (solid colored lines).}
  \label{fig:analytical}
\end{figure}

The next step is to subtract the right hand side of \eqref{eq:step1} from unity, to get the probability that a given minority node does \emph{not} vote. Since there are $N_{-}$  such nodes, and their decisions to vote are all independent, the probability that all of them do not vote is: 
 \begin{equation}
 \begin{aligned}
     P(\text{no minority nodes vote}) & = \\ \left[1-P(\text{any given minority node votes})\right]^{N_{-}}.
\end{aligned}
 \end{equation}
Then, by subtracting this quantity from $1$, we obtain the probability that at least one minority node votes,
 \begin{equation}
 \begin{aligned}
     P(\text{at least one minority node votes}) &= \\ 1-P(\text{no minority nodes vote}).
     \label{eq:analytical_expression}
\end{aligned}
 \end{equation}
As stated above, this probability is also equal to the probability that the minority wins. Combining the equations above and  replacing $N_{-}$ with $\alpha N_{+}$ throughout, we finally arrive at our desired result: 
  \begin{equation}
 \begin{aligned}
     & P(\text{minority wins}) \\ &= 1- \Big[1-{N_{+} \choose {\alpha N_{+}}} p_{\text{out}}^{\alpha N_{+}} \left(1-p_{\text{out}}\right)^{N_{+}-\alpha N_{+}} \\
     &-{N_{+} \choose \alpha N_{+}-1} p_{\text{out}}^{\alpha N_{+}-1} \left(1-p_{\text{out}}\right)^{N_{+}-(\alpha N_{+}-1)}\Big]^{\alpha N_{+}}.
     \end{aligned}
     \label{eq:all-alpha-maintext}
 \end{equation}
Figure \ref{fig:analytical} shows an excellent match between this analytical prediction and simulations.
 
\subsubsection{Peak location and probability of a minority victory} 
In Figure~\ref{fig:SB} we saw that the probability of the minority winning in our simulations on stochastic block networks was reached at high $p_{\rm in}$ and intermediate values of $p_{\rm out}$. Continuing to assume fully connected blocks, $p_{\rm in}=1$, we can now calculate at the value of $p_{\rm out}$ that maximizes the probability of a minority victory. To do so, we differentiate Eq.~\eqref{eq:all-alpha-maintext} with respect to $p_{\text{out}}$ and set the resulting expression to zero.  
After straightforward but extensive algebra, and with the help of Stirling's formula, we find that in the limit $N_+ \rightarrow \infty$ with $\alpha$ held fixed, $$p_{\text{out}}=\alpha$$ 
maximizes the probability of a minority victory. 

\begin{figure}[htb]
    \centering
    \includegraphics[width=\linewidth]{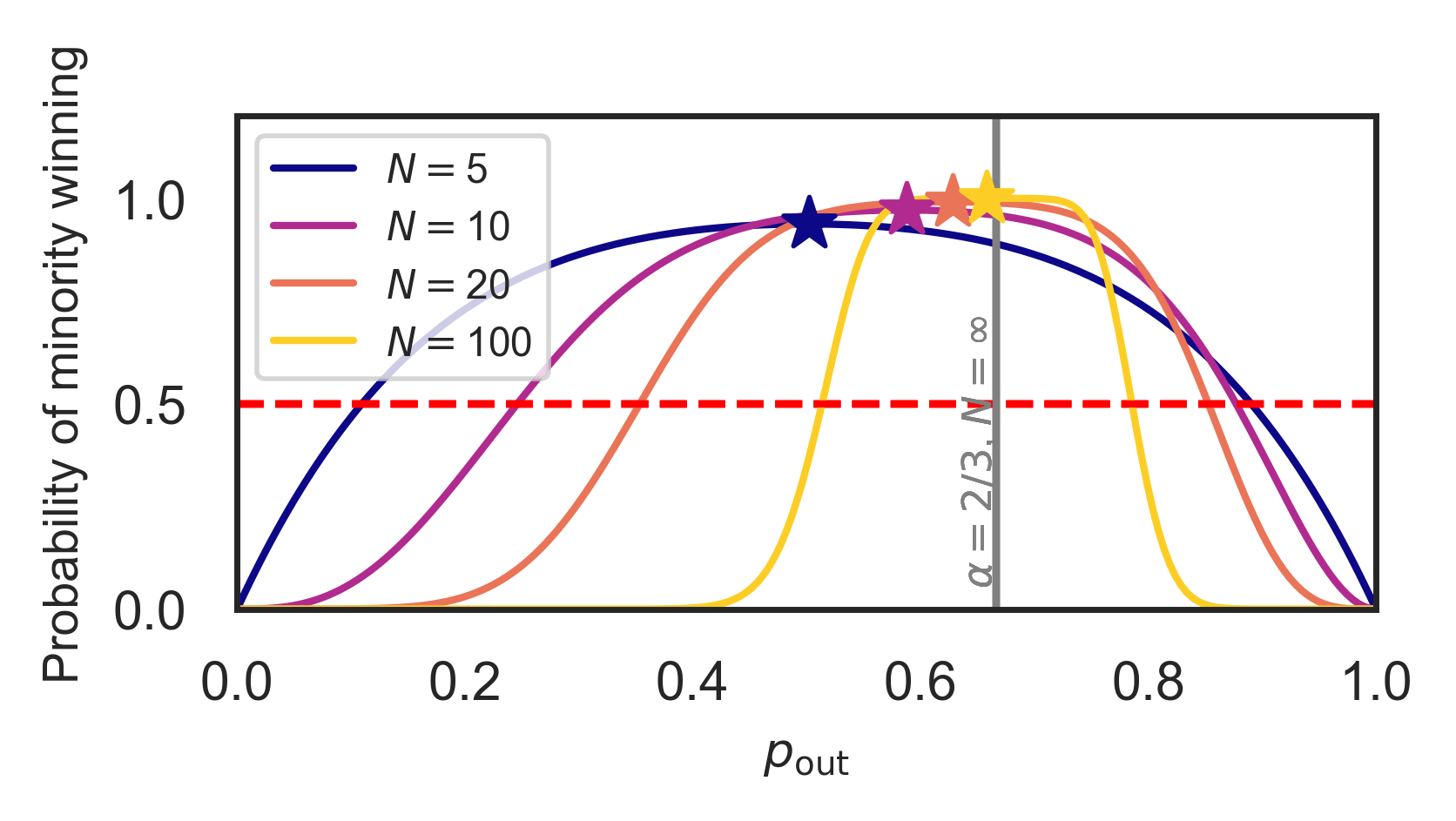}
    \caption{ The solid curves show proportion of undemocratic outcomes as described by Eq.~\eqref{eq:all-alpha-maintext} for constant $\alpha=2/3$ and $N=5, \ 10, \ 20, \ 100$. The stars indicate the location of the maximum on each curve. The probability $p_{\rm out}$ that maximizes the proportion of undemocratic outcomes approaches $\alpha$ as $N$ increases. The limiting location of the peak, $p_{\rm out}=\alpha$, is marked by the gray vertical line.} 
  \label{fig:alpha_convergence}
\end{figure}

Figure \ref{fig:alpha_convergence} shows how the peak value of $p_{\text{out}}$ converges to $\alpha$ as $N$ increases. In these plots, we fix $\alpha=N_{-}/N_{+}=2/3$ and vary the network size $N$. Notice that at the peak, the proportion of undemocratic outcomes  approaches 1 as $N$ goes to infinity. With further effort, one can show that the peak value of a minority victory deviates from 1 by an exponentially small term for $N \gg 1$: 
 \begin{equation}
  \begin{aligned}
  P(\text{minority wins} \,|\, p_{\text{out}}=\alpha) \sim
 & 1- \exp\left(-\sqrt{\frac{2 \alpha N }{\pi (1-\alpha^2)}} \, \right) \\
 &\times  \left(\exp\left[-\frac{1}{(1-\alpha) \pi}\right]\right).
 \end{aligned}
 \end{equation}

Furthermore, the curves in Fig.~\ref{fig:alpha_convergence} become increasingly sharply peaked as $N$ increases. To check this, we evaluate Eq.~\eqref{eq:all-alpha-maintext} in the same way at $p_{\text{out}}=\alpha+\epsilon$ for $\epsilon \ll 1$ and find that $P(\text{minority wins}\,|\, p_{\text{out}}=\alpha+\epsilon)$ tends to 0 as $N$ approaches infinity. Therefore in the large-$N$ limit, $P(\text{minority wins})$ tends to a discontinuous function that equals 1 at $p_{\text{out}}=\alpha$ and 0 everywhere else.

\subsection{Networks with a heavy-tailed degree distribution}
\label{sec:scale-free}
\ER networks and stochastic block networks are both widely studied. 
Their simplicity allowed us to derive analytical results and gain some intuition for when the minority could win the election in our model. 
In both models, however, nodes tend to have very similar numbers of network neighbors. 
This homogeneity is different from many real-world networks in which node degrees can vary a lot~\cite{,barabasi1999emergence,newman_networks_2018,clauset2009power}.

As an example of networks with broad degree distributions we now consider networks whose degree distributions follow a power law in the limit $N\to \infty$. Such scale-free networks have been claimed to capture features of many real-world networks~\cite{albert2002statistical, barabasi1999emergence}. Other scholars have moderated or even argued against this claim~\cite{clauset2009power,broido2019scale}. 

In our investigations of stochastic block networks, we found that the existence of community structure could in some cases increase the likelihood of a minority win under our model. To understand the effect of homophily in more detail, we also incorporate homophily in our simulations of our model in networks with a heavy-tailed degree distribution. In order to introduce homophily into the setting of networks with power-law degree distributions, we introduce a homophily parameter $h$ ($0\leq h \leq 1$). When $h=0$, the node opinions are distributed randomly on the network, whereas when $h=1$, the majority and minority nodes organize into disjoint blocks with no connection between nodes of different opinions. Our algorithm for generating homophily on networks with power-law degree distributions is described in Appendix \ref{sec:algorithm}. The algorithm is heavily inspired by algorithms used to create configuration-model networks~\cite{newman_networks_2018}. In that sense, our networks with power-law degree distributions can be thought of as a class of configuration-model networks with homophily. 

\begin{figure}[!h]
  \centering
         \includegraphics[width=0.4\linewidth]{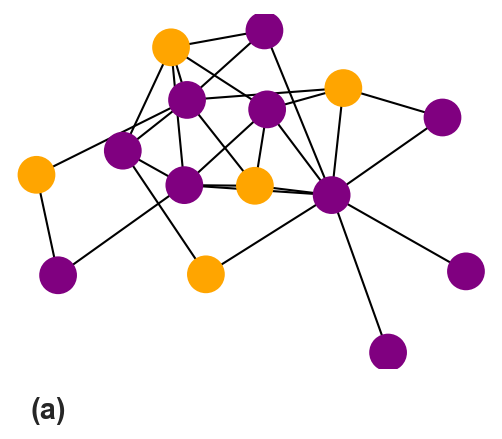} 
        \includegraphics[width=0.4\linewidth]{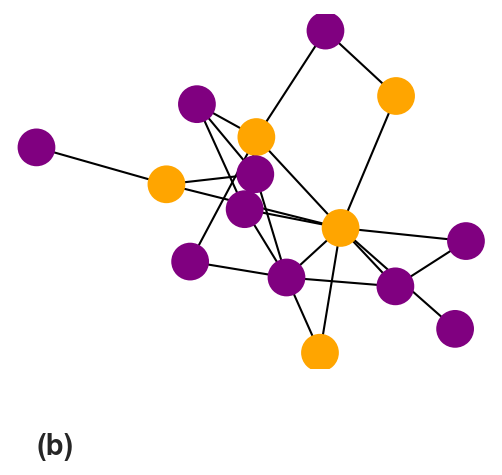} \\
        \includegraphics[width=0.4\linewidth]{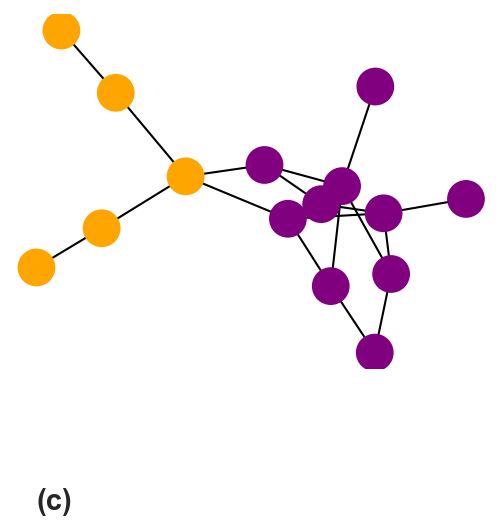} 
         \includegraphics[width=0.4\linewidth]{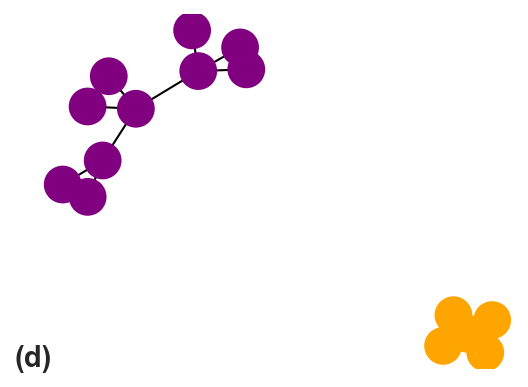}        

         \caption{Examples of networks with heavy-tailed degree distributions with homophily factors (a) $h=0$ , (b) $h=0.3$, (c) $h=0.8$, and (d) $h=1$. All networks are of size $N=15$ with minority size $N_{-}=5$. We use the power law exponent $\lambda=2.5$ to generate the degree distribution.}
         \label{fig:scale_free_example}
\end{figure}

Figure \ref{fig:scale_free_example} shows examples of the resulting networks. As $h$ increases, the nodes get a higher preference for connecting to nodes with the same opinion. When $h=0$, the nodes' local information is most likely to be representative of the true proportion of opinions across the electorate as a whole. When $h=1$, the nodes' local information will only reflect the presence of nodes with the same opinion. 

\begin{figure}[!t]
  \centering

      \includegraphics{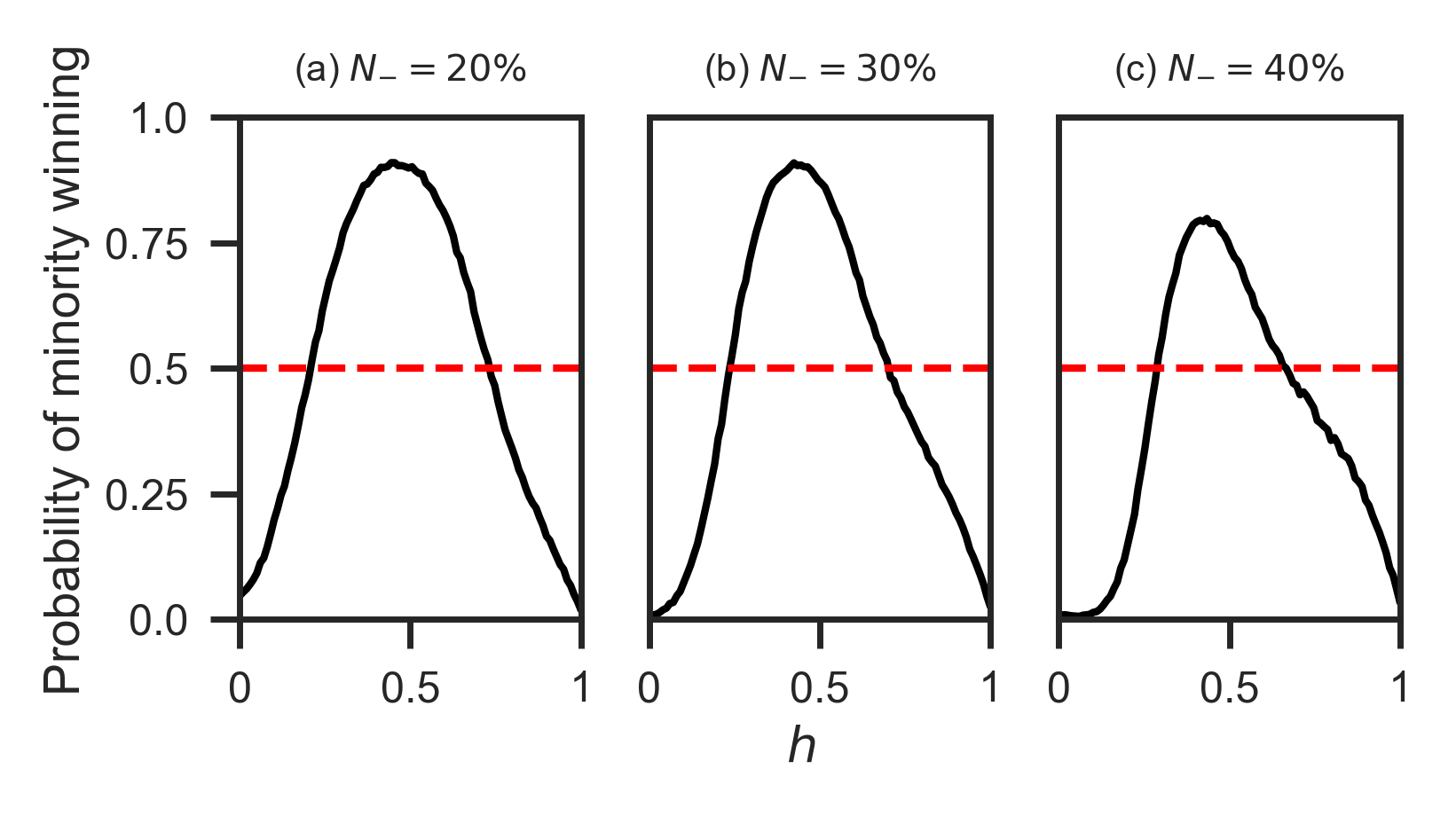}

         \caption{Proportion of minority winning on networks with heavy-tailed degree distributions and size $N=100$, for (a) $N_{-}$ = 20, (b) 30, and (c) 40 nodes as a function of the homophily factor $h$, for $10^4$ simulations.}
         \label{fig:SF_proportion}
\end{figure}

Figure~\ref{fig:SF_proportion} shows the proportion of undemocratic outcomes on a network with heavy-tailed degree distribution and size $100$ with minority fractions $20 \%$, $30\%$, and $40\%$. The horizontal axis shows the homophily parameter $h$. We observe once again that undemocratic outcomes occur most frequently when the homophily parameter is in the intermediate range.  In Figs.~\ref{fig:SF_proportion}(a) and (b), for  homophily parameter values in range $0.3\leq h \leq 0.7$ the minority faction wins more than half of the time. Surprisingly, increasing the minority size $N_{-}$ does not yield a larger peak probability of minority wins for these configuration networks, in contrast to the other network structures tested in this paper.

\begin{figure}
      \includegraphics{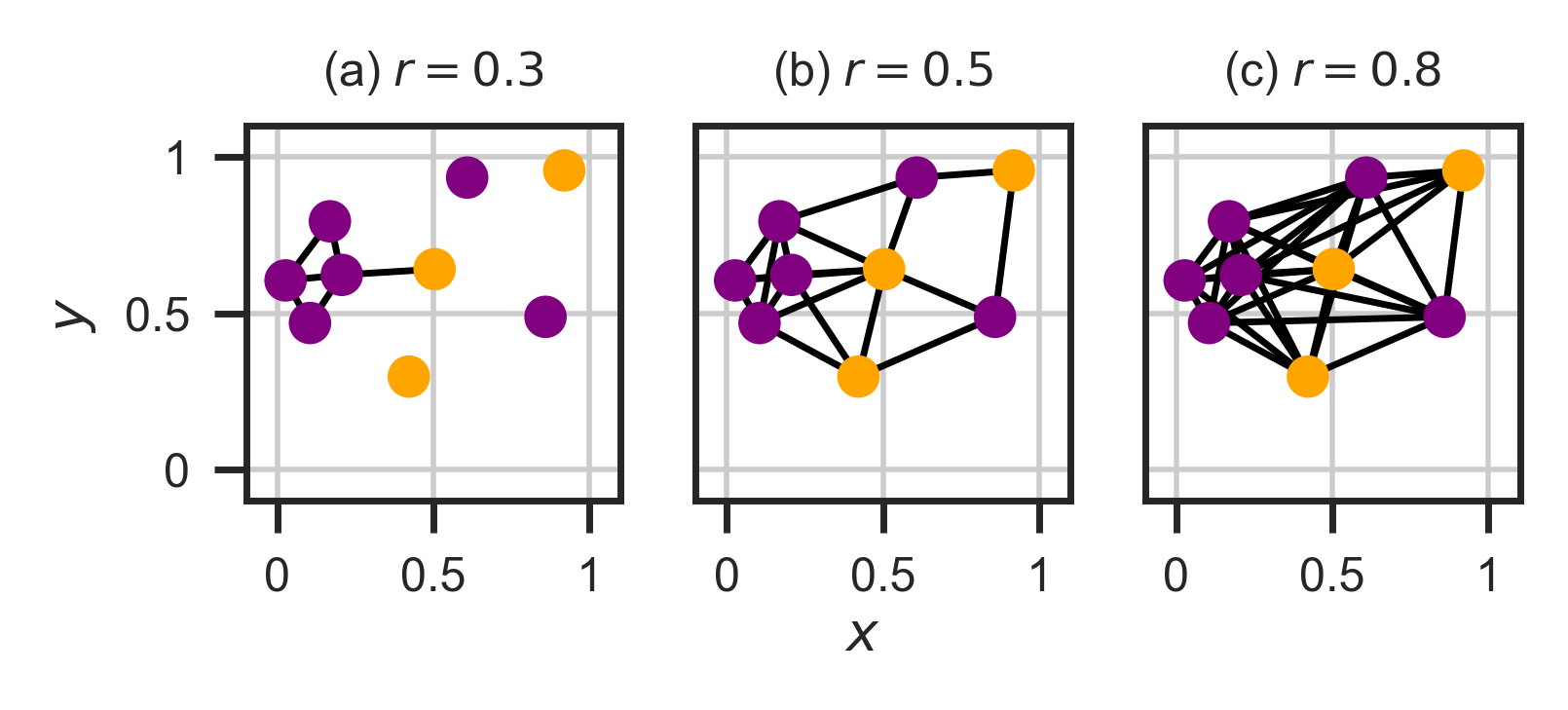}
         \caption{Example of majority (orange) and minority (purple) node distributions for geometric random networks with radius (a) $r=0.3$, (b) $r=0.5$, and (c) $r=0.8$.}
         \label{fig:geo_example}
\end{figure}

\subsection{Geometric Random Networks}
\label{sec:geo-rand}
In Section \ref{sec:scale-free}, we considered networks with broad degree distributions, a trait shared by some social networks. A qualitatively different class of networks are those in which the likelihood of a link between two nodes depends on their geographical separation. ``Geometric random networks'' provide some of the simplest examples. To generate them, imagine throwing nodes uniformly at random inside a unit square. We add an edge between any two nodes that lie within a distance $r$ of each other. A larger value of $r$ results in denser networks, as illustrated in Fig.~\ref{fig:geo_example}. 

In order to incorporate homophily into these sorts of random networks, we assign minority and majority opinions preferentially to the left and right halves of the unit square, respectively. With probability equal to the homophily parameter $h$, nodes lie within their preferred half of the square.

\begin{figure}[!h]
  \centering

        \includegraphics{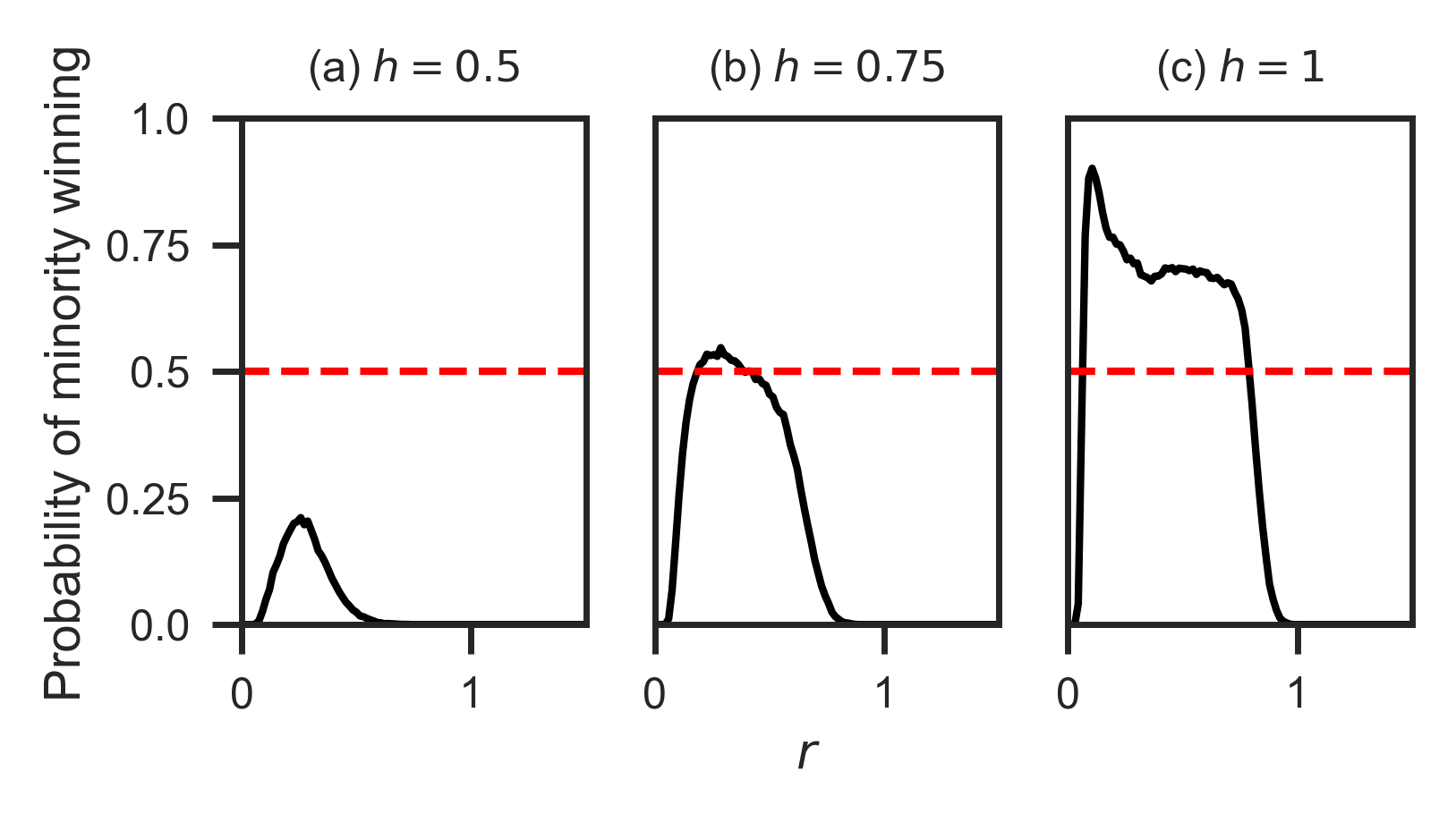}
         \caption{Proportion of minority victories on a geographic random network as a function of the radius of connection $r$ and probability of nodes lying within their preferred half of the square. (a) $h=0.5$ (no homophily), (b) $h=0.75$ (moderate homophily), and (c) $h=1$ (extreme homophily). The results are for networks of size $N=100$, with minority size $N_{-}=30$, for $10^4$ simulations.}
         \label{fig:geo_proportion}
\end{figure}

We vary the radius of connection $r$ and compute the proportion of undemocratic outcomes. Figure \ref{fig:geo_proportion} shows the results of the simulation. While the proportion of undemocratic outcomes peaks in the intermediate radius range, the peak probability of minority victories  moves to the right as homophily increases. In a low-homophily setting, minority nodes benefit from low radius to prevent dejectedness (they need to actively avoid knowing majority nodes). In a high-homophily setting, minority nodes benefit from a higher radius to prevent complacency (they need to ensure they know some majority nodes). The extreme peak in panel (c) is interesting. It is due to the fact that in extreme homophily settings, the majority half of the square domain is more densely populated. Therefore, at low non-zero values of $r$, majority nodes begin to see other majority nodes and become complacent before minority nodes begin to see other nodes. This effect results in many disconnected minority nodes voting. The effect is diminished when the difference between $N_{+}$ and $N_{-}$ is lower. While not shown here, our numerical experiments show that the peak is higher for $N_{-}=20\%$ and lower for $N_{-}=40\%$.

\section{Discussion}
\label{sec:discussion}
In this paper, we have presented a simple agent-based model of voter turnout. 
By simulating the model on a variety of network structures, we found that it is often possible for a minority faction to win the model election under the effects of dejectedness and complacency. 
These undemocratic outcomes occur most frequently in the parameter ranges that correspond to intermediate knowledge of the global state of the electorate, as well as in networks with some homophily or community structure.
We have further shown that undemocratic outcomes can become more likely in settings where the local distribution of opinions is not representative of the average global distributions. Intermediate homophily settings often create regimes in which minority nodes are more likely to overestimate the closeness of an election based on their one-hop network neighborhood, while majority nodes are susceptible to complacency.
 
In reality, it remains unknown how much complacency and dejectedness influence whether people cast their vote in elections. It is also unknown to what extent such complacency and dejectedness would be caused by the immediate social-network neighborhood of the voter; it seems quite possible, for example, that media reports, forecasting agencies, and other non-local effects could play even bigger roles in pushing voters to turn out or stay home. All that one can say with certainty is that voter decision-making is a complex phenomenon with many social, political, and structural factors influencing individual choices. 
Nonetheless, our work suggests that homophily and network structure can greatly affect vote outcomes in settings where voters choose to abstain or cast their ballots based on the prevalence of opinions in their local social neighborhood.

There are many extensions of this study that would be intriguing to try in future work.
Some directions could focus on implementing the model in more general settings such as: 
Realizing the model on core/periphery networks, adding more than two opinion states, modifying the decision rule, and perhaps adding a tension between local and global information in the form of broadcasters or forecasters.
Another possibility would be to make the model dynamic. 
What do nodes do after having lost an election that they thought was a safe win? 
Introducing such dynamics and looking for fixed points, cycles, and other time-varying states would be interesting.

\section{Acknowledgements}
We thank Moon Duchin and Stephen Cowpar for helpful advice and discussions. Research supported by a Cornell Center of Applied Mathematics postdoctoral fellowship (J.L.J.) and NSF grant CCF-1522054 and National Institutes of Health grant 1R37CA244613-01 (S.H.S.).

\bibliography{pre-submission-1}

\appendix

\section{Algorithm for generating scale-free networks with homophily}
\label{sec:algorithm}

We generate a scale-free network with homophily using the following algorithm:
\begin{enumerate}
    \item Fix $n$ nodes.
    \item Draw degrees from a power law distribution. 
    \item Generate a vector of length $n$ assigning a binary opinion: 0 to majority nodes and 1 to minority nodes.
    \item Initialize two empty stacks: the majority stack and the minority stack.
    \item For each node: \\
    If a node is a minority node, add its index to the minority stack the number of times corresponding to its degree. \\
    If the node is a majority node, add its index to the majority stack the number of times corresponding to its degree.
    \item Shuffle the majority and minority stacks.
    \item While the minority stack is non-empty: 
    \\ pop \textbf{node1} from the top of the minority stack. generate a random number between 0 and 1. If the random number is less than the homophily factor $h$, draw an edge between \textbf{node1} and the first node in the minority stack (\textbf{node2}). If the random number is greater, draw an edge between node1 and the top node in the majority stack.
    \item If the majority stack is nonempty by the time the minority stack is empty, connect the remaining majority nodes in pairs. 
\end{enumerate}

\end{document}